\documentstyle[psfig]{mn}

\title[Late-type stars members of young stellar kinematic groups]
{Late-type stars members of young stellar kinematic groups --
 I. Single stars}
\author[D. Montes et al.]
  {D.~Montes,\thanks{E-mail: dmg@astrax.fis.ucm.es} 
J. L\'{o}pez-Santiago,
M.C. G\'{a}lvez,
M.J. Fern\'{a}ndez-Figueroa, 
\newauthor E. De Castro, M.~Cornide\\
  Departamento de Astrof\'{\i}sica, Facultad de F\'{\i}sicas,
  Universidad Complutense de Madrid, E-28040 Madrid, Spain\\
}
\date{Accepted 2001 ... . Received 2001  ... }
\pagerange{\pageref{firstpage}--\pageref{lastpage}}
\pubyear{2001}

\def\LaTeX{L\kern-.36em\raise.3ex\hbox{a}\kern-.15em
    T\kern-.1667em\lower.7ex\hbox{E}\kern-.125emX}

\begin{document}

\label{firstpage}

\maketitle

\begin{abstract}
This is the first paper of a series aimed at studying
the properties of late-type stars members of young stellar kinematic groups.
We concentrate our study on classical young moving groups as:
Local Association (Pleiades moving group, 20 - 150 Myr),
IC 2391 supercluster (35 Myr),
Ursa Major group (Sirius supercluster, 300 Myr),
 and  Hyades supercluster (600 Myr);
as well as on recently identified groups as: 
Castor moving group (200 Myr).
In this paper we have compiled a preliminary list 
of single late-type stars
possible members of some of these young stellar kinematic groups.
Stars have been selected from previously established members of
stellar kinematic groups based on photometric and kinematic properties
as well as from candidates based on other criteria as their
level of chromospheric activity, rotation rate, lithium abundance.
Precise measurements of proper motions and parallaxes taken from
Hipparcos Catalogue, as well as from Tycho-2 Catalogue, 
and published radial velocity measurements
are used to calculate the Galactic space motions (U, V, W)
and to apply the Eggen's kinematic criteria
in order to determine the membership of the selected stars to the 
different groups.
Additional criteria using age-dating methods for late-type stars
will be applied in forthcoming papers of this series.
A further study of the list of stars compiled here 
could lead to a better understanding of the chromospheric activity 
and their age evolution as well as of the 
star formation history in the solar neighbourhood.
In addition, these stars are also potential search targets 
for direct imaging detection of sub-stellar companions

\end{abstract}


\section{Introduction}

Stellar kinematic groups (SKG) 
are kinematically coherent groups of stars
that could share a common origin (the evaporation of an open cluster, 
the remnants of a star formation region
or a juxtaposition of several little star formation bursts at different 
epochs in adjacent cells of the velocity field).
Eggen (1994) defined a ``supercluster" (SC) as a group of stars
gravitationally unbound that share the same kinematics and may
occupy extended regions in the Galaxy, and a ``moving group" (MG)
as the part of the supercluster that enters the solar neighbourhood
and can be observed  all over the sky.
It has long been known that in the solar vicinity there are several 
groups of stars that share the same space motions than well known
open clusters.
The youngest and best documented groups are the 
Hyades supercluster (Eggen 1958a, 1960a, 1984a, 1992b, 1996, 1998b) 
associated with the Hyades cluster (600 Myr),
and the Ursa Major group (Sirius supercluster) 
(Eggen 1960b, 1983a, 1992a, 1998c, Soderblom \& Mayor 1993a, b) 
associated with the UMa cluster of stars (300 Myr).
A younger kinematic group called the Local Association or Pleiades moving group seems to consist of a reasonably
coherent kinematic stream of young stars with embedded clusters 
and associations such as the Pleiades, $\alpha$ Per,
NGC 2516, IC 2602, and Scorpius-Centaurus  
(Eggen 1975, 1983b, c, 1992c, 1995a).
The age of the stars of this association ranges from about 20 to 150 Myr.
Evidences have been found that X-ray and EUV selected active stars and 
lithium-rich stars (Favata et al. 1993, 1995, 1998; 
Jeffries \& Jewell 1993; Mullis \& Bopp 1994; Jeffries 1995)
are members of this association.
Other two young moving groups are the 
IC 2391 supercluster (35-55 Myr) (Eggen 1991, 1995b) and
the Castor Moving Group (200 Myr) (Barrado y Navascu\'es 1998).

\begin{table*}
\caption[]{Young stellar kinematic groups
\label{tab:skg}}
\begin{flushleft}
\small
\begin{tabular}{llcccccccccccccccccccccccc}
\noalign{\smallskip}
\hline
\noalign{\smallskip}
Name & Cluster(s) & Age   & U, V, W     &  V$_{\rm T}$  &  C.P. (A, D)  \\
     &           & (Myr) & (km s$^{-1}$) &  (km s$^{-1}$)  & ($^{h}$, $^{o}$)    \\
\noalign{\smallskip}
\hline
\noalign{\smallskip}
Local Association & Pleiades, $\alpha$ Per, M34 & 20 -- 150 &
 $-11.6, -21.0, -11.4$ & $26.5$ & $(5.98, -35.15)$ \\
(Pleiades moving group) & $\delta$ Lyr, NGC 2516, IC2602,      &          &
                     &      &               \\
\noalign{\smallskip}
IC 2391 supercluster & IC 2391          & 35 -- 55 & 
 $-20.6,  -15.7,  -9.1$  & $27.4$  & $(5.82, -12.44)$  \\
\noalign{\smallskip}
Castor moving group &                  & 200  & 
 $-10.7,  -8.0,  -9.7$   & $16.5$ & $(4.75, -18.44)$ \\
\noalign{\smallskip}
Ursa Major group    & Ursa Major       & 300  &  
$14.9, 1.0, -10.7$  & $18.4$ & $(20.55, -38.10)$ \\
(Sirius supercluster) &      &          &
                     &      &               \\
\noalign{\smallskip}
Hyades supercluster & Hyades, Praesepe & 600  & 
$-39.7, -17.7, -2.4$ & $43.5$ & $(6.40, 6.50)$  \\
\noalign{\smallskip}
\hline
\end{tabular}

\end{flushleft}
\end{table*}

Since  Olin Eggen introduced the concept of MG and
the idea that stars can
maintain a kinematic signature over long periods of time, their
existence (mainly the old MGs) has been rather controversial 
(see Griffin 1998; Taylor 2000). 
There are two factors that act against the persistence of a MG:
the Galactic differential rotation (tends to spread the stars) and
the disc heating (velocity dispersion of disc stars increase with age).
However, recent studies (Dehnen 1998; 
Chereul, Cr\'ez\'e \& Bienaym\'e 1998, 1999; 
Skuljan, Hearnshaw \& Cottrell 1999; 
Asiain et al. 1999; 
Torra, Fern\'andez \& Figueras 2000;
Myll\"ari, Flynn \& Orlov 2000; 
Feltzing \& Holmberg 2000)
using astrometric data taken from Hipparcos and different procedures 
to detect MG not only confirm the existence of classical young MGs
(and some old MGs),
but also detect finer structures in space velocity and age
that in several cases can be related to kinematic properties of nearby 
open clusters or associations.
Skuljan, Cottrell \& Hearnshaw (1997) have also confirmed the Eggen's 
hypothesis of MG using Hipparcos astrometric data.
These authors found that the use of Hipparcos data considerably improve 
the velocity dispersions for all the Eggen's MGs.
However, the Eggen's membership criterion of constant V
is not confirmed and they conclude that both U and V velocity 
components must be used to create more realistic membership criteria.
More complex structures characterized by several longer {\em branches} 
(Sirius, middle, and Pleiades branches)
running almost parallel to each other across the UV-plane 
have been found by Skuljan et al. (1999) 
in their study of the velocity distribution of stars in the solar 
neighborhood.

Well known members of these moving groups are mainly early-type stars
and few studies have been concentrated on late-type stars.
However, evidences have been found that many young late-type stars
can be members of some young MG
(X-ray and EUV selected active stars and lithium-rich
stars (Jeffries 1995); the late-type stellar population of the Gould belt
(Guillout et al. 1998; Makarov \& Urban 2000)).
Identification of a significant number of 
late-type members of these
young moving groups would be extremely important for a study of the
chromospheric and coronal activity and their age evolution.
This is the aim of  this series of papers.

In this first paper we focus on the compilation of a preliminary list 
of single late-type stars,
previously established members
or possible new candidates of the different young SKG
 mentioned above (see Table~\ref{tab:skg}).
We have examined the kinematic properties of these stars
 using the more recent radial velocities and astrometric data available,
in order to determine their membership to the different SKG.
In a companion paper (Montes et al. 2001c; hereafter Paper II) we give the list
 of spectroscopic binaries, some of them well known chromospherically 
active binaries (for preliminary results see 
Montes, Latorre \& Fern\'andez-Figueroa 2000a; and Montes et al. 2001a).
The origin of these young SKG will be addressed in Paper III.
With this aim we have taken the most recent
data available in the literature (including astrometric data from
Hipparcos Catalogue and new Tycho-2 Catalogue) of the nearby young
open clusters, OB associations,
T associations, and other associations of young stars as TW Hya,
in order to calculate their Galactic space motions (U, V, W)
and space coordinates (X, Y, Z) and to study their possible association
with the different young SKG as well as with
the young flattened and inclined Galactic structure known as the 
Gould Belt (for preliminary results see Montes 2001a, 2001b;
for a review about the evolution from OB associations and moving groups 
to the field population see Brown 2001).

In addition to the kinematic properties we have also compiled for each star
the photometric, spectroscopic and physical properties as well as
information about activity indicators and Li abundance.
For some of the candidate stars included in the list analysed in 
this paper we have also taken high resolution
echelle spectra in order to obtain
a better determination of their radial velocity, lithium abundance,
 rotational velocity and the level of chromospheric activity 
(for preliminary results see  Montes 2001b; 
Montes, L\'opez-Santiago \& G\'alvez 2001b; Montes et al. 2001d).
We will use all this data in forthcoming papers to analyse in detail
the membership to the different young SKG and identified possible
age subgroups (see Barrado y Navascu\'es 1998, 
Barrado y Navascu\'es et al. 1999; Song et al. 2000;
L\'opez-Santiago, Montes \& G\'alvez 2001).

In Sect. 2 we describe the young SKG we have considered in this work.
The details of the sample selection are given in Sect. 3.
In Sect. 4 we analyse the membership of this sample to the different
SKG using as membership criteria the 
Galactic space-velocity components (U, V, W)
 and the Eggen's kinematic criteria. 
Finally, Sect. 5 gives the discussion and conclusions.

\begin{figure*}
{\psfig{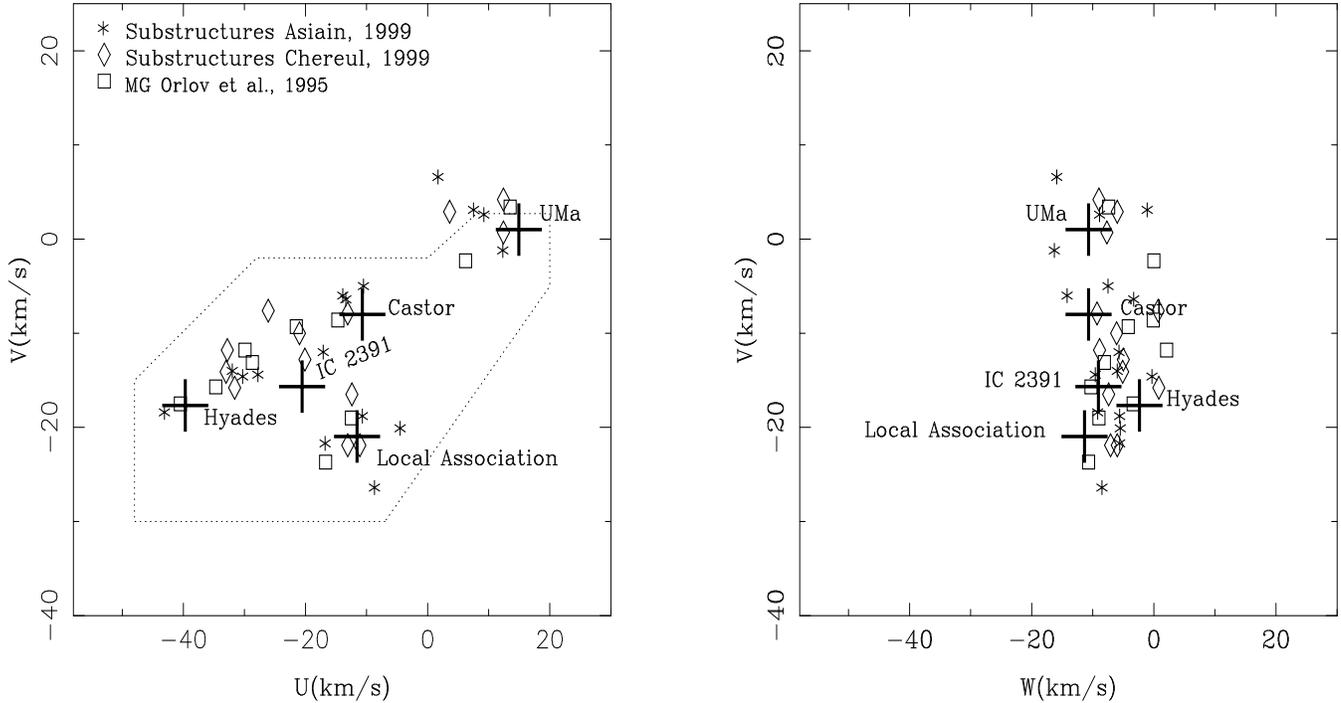}}
\caption[ ]{
(U, V) and (W, V) planes (Boettlinger Diagram) 
in the region of the young disk stars.
Big crosses are centered at the five young SKG
analysed in this work, as given in Table~\ref{tab:skg}.
Different symbols are used to plot the position of other SKG and 
substructures found by other authors.
The dashed line represents the  boundaries that determine the young disk
population as defined by Eggen (1984b, 1989).
\label{fig1} }
\end{figure*}

\section{Young SKG}

We focus our study here on the five youngest and best documented SKG:
the Local Association (Pleiades moving group, 20 -- 150 Myr),
IC 2391 supercluster (35 Myr),
Castor moving group (200 Myr),
Ursa Major group (Sirius supercluster, 300 Myr),
 and  Hyades supercluster (600 Myr).
The properties of these SKG are summarized in Table~\ref{tab:skg}.
We list the name, possible open clusters associated to the group,
range of age (Myr), the Galactic space-velocity components
(U, V, W), total velocity (V$_{\rm T}$) and the coordinates (A, D) of the
convergent point (C.P.).
The velocity vector have been calculated by us using the 
spherical parameters and V$_{\rm T}$ assigned to each group in the literature
(Eggen 1958b, 1984c, 1991, 1992c). 
For the recently identified Castor MG the
C.P. and V$_{\rm T}$ have been derived by us from the space-velocity components
given by (Barrado y Navascu\'es 1998).
In Fig.~\ref{fig1} we have plotted the position of these SKG in the 
 (U, V) and (W, V) planes.
The velocity components of the substructures found in these SKG by
Asiain et al. (1999) using statistical methods 
and Chereul et al. (1999) using a 3-D wavelet analysis both in 
the density and velocity distributions are also plotted in 
Fig.~\ref{fig1}.
Orlov et al. (1995) using a hierarchical clustering method 
have found several kinematic
groups in the solar neighborhood which  velocity components 
are close to the five young SKG we considered here.
We have plotted the U, V, W components of these MG in  Fig.~\ref{fig1} 
for comparison.

\section{Selection of the Sample}

The sample of late-type stars (spectral type later than F2) 
analysed in this work 
has been selected from previously 
established members of some of these SKG, based on the photometric 
and kinematic properties as well as from new candidates based on other 
criteria as their level of chromospheric and coronal activity, 
rotation rate, and lithium abundance, which are spectroscopic signatures
of youth.

On the one hand the rotation rate in late-type stars
moderates the dynamo
mechanism which generates and amplifies the magnetic fields in the
convection zone, but there is a further relationship between 
rotation and age.
Rotation rates decline with age because stars lose angular momentum
through the coupling of the magnetic field and stellar mass loss,
and thus there is an indirect trend of decreasing magnetic activity 
with increasing age.
On the other hand
the resonance doublet of  Li~{\sc i} at $\lambda$6708 \AA\
is an important diagnostic of age in late-type stars
since it is easy destroyed by thermonuclear reactions in the
stellar interiors.
Therefore, high level of magnetic activity, 
rapid rotation, and strong lithium absorption are 
spectroscopic signatures of youth, and the stars selected in this way
are good candidates to be members of some of the young SKG we are 
analysing here.

The main sources from which we have selected this late-type star sample are:

\begin{itemize}

\item The membership lists given by Eggen in his four decades
of research on SKG (Eggen 1958 to 1998), and additional lists given by
Soderblom \& Mayor (1993a).

\item The study of Agekyan \& Orlov (1984) and Orlov et al. (1995) 
which searched for kinematic 
groups in the solar neighborhood 
(see also Popovi\'c, Ninkovi\'c \& Pavlovi\'c  1995).

\item The study of ages of spotted late-type stars by Chugainov (1991).

\item X-ray and EUV selected active stars and lithium-rich stars
(Favata et al. 1993, 1995, 1998; Jeffries \& Jewell 1993; 
Tagliaferri et al. 1994; Mullis \& Bopp 1994;
Jeffries 1995; Schachter et al. 1996; 
H\"unsch, Schmitt \& Voges 1998a, b; H\"unsch et al. 1999; 
Cutispoto et al. 1999, 2000).

\item Single rapidly rotating stars as AB Dor, PZ Tel, HD 197890, 
RE J1816+541, BD+22 4409 (LO Peg), HK Aqr,
V838 Cen, V343 Nor, LQ Hya,
previously assigned membership of the Local Association.

\item Chromospherically active late-type dwarfs in the solar 
neighborhood with studied kinematic properties
(Young, Sadjadi \& Harlan  1987; Upgren 1988; Soderblom \& Clements 1987; 
Soderblom 1990; Ambruster et al. 1998).

\item Flare stars with studied kinematic properties (Poveda et al. 1996).

\item The study of field M dwarfs with high resolution spectra by
Delfosse et al. (1998), including the recently identified M9V star 
DENIS 1048-39 which is the closest star later than M7V 
(Delfosse et al. 2001).

\item Other chromospherically active stars  
(Henry, Fekel \& Hall 1995; Henry et al. 1996; Soderblom et al. 1998).

\item Late-type stars included in the list of the nearest 100 Stellar Systems
given by the Research Consortium on Nearby Stars (RECONS\footnote{
 RECONS: {\tt http://joy.chara.gsu.edu/RECONS/}
}).

\item The study of nearby young solar analogs by Gaidos (1998) and
Gaidos, Henry \& Henry (2000).

\item The sample of nearby, single, solar-type stars
selected as proxies for the Sun at different stages in the project 
the ``Sun in Time" by Bochanski et al. (2000).

\item The study of nearby young x-ray active low-mass stars with
well-measured parallaxes by Wichmann \& Schmitt (2001).

\item The active stars included in the the Vienna-KPNO search for 
Doppler-imaging candidate stars (Strassmeier et al. 2000).

\end {itemize}

For this selected sample we only analysed here single stars 
or effective single stars (wide visual binaries). The spectroscopic binaries 
are analysed in Paper II.
We have considered only isolated stars, that is, 
we have excluded from the sample known members of open clusters, 
OB associations.
However, we have included some members of 
other associations of young stars as TW Hya
and the recently identified
$\beta$ Pic moving group (Barrado et al. 1999);  
Tucanae association (Zuckerman \& Webb 2000);  
Horologium association (Torres et al. 2000); and
HD 199143 stellar group (van den Ancker et al. 2000; 
van den Ancker, Pérez \& de Winter 2001),
which could be stream stars related with the Local Association
(see Montes 2001a, b; L\'opez-Santiago, et al. 2001).

Some pre-main sequence late-type stars (weak T Tauri stars (WTTS), and
post T Tauri stars (PTTS) are also included in our sample as possible 
members of the youngest SKG, the Local Association.
Oppenheimer et al. (1997) have identified two very young M dwarfs
which also could be members of the Local Association.
In the last years many WTTS, PTTS and young zero-age main sequence 
stars have been identified 
(using Li as an age criterion) with optical
follow-up spectroscopy of ROSAT X-ray sources 
in and around nearby star-forming regions and OB associations.
Some of these stars could be members of the Local Association
as has been suggested by Mart\'{\i}n \& Magazz\`u (1999) and Frink (2001)
or may represent a population of Gould Belt low-mass stars 
(Wichmann et al. 1999, 2000).
We have not included these new identified young stars in our sample, 
because only a few have enough data (astrometric data and radial velocities)
to analyse their kinematic, but they will be included in future work.

Our sample also includes some of the host stars of extra-solar planets
discovered in the past few years by measuring their Keplerian Doppler shifts 
(up today more than 60, see Marcy, Cochran \& Mayor 2000; 
Marcy \& Butler 2000).
These stars are nearby late-type stars
with high precision radial velocity measurements.
Although, many of them have ages greater than 3~Gyr as has been derived 
using evolutionary tracks (Fuhrmann, Pfeiffer \& Bernkopf 1998; 
Ford, Rasio \& Sills 1999)
or Ca~{\sc ii} H\&K fluxes (Henry et al. 2000), 
others are known to be younger and then could be possible members 
of some of the young SKG analysed here.

\begin{figure*}
{\psfig{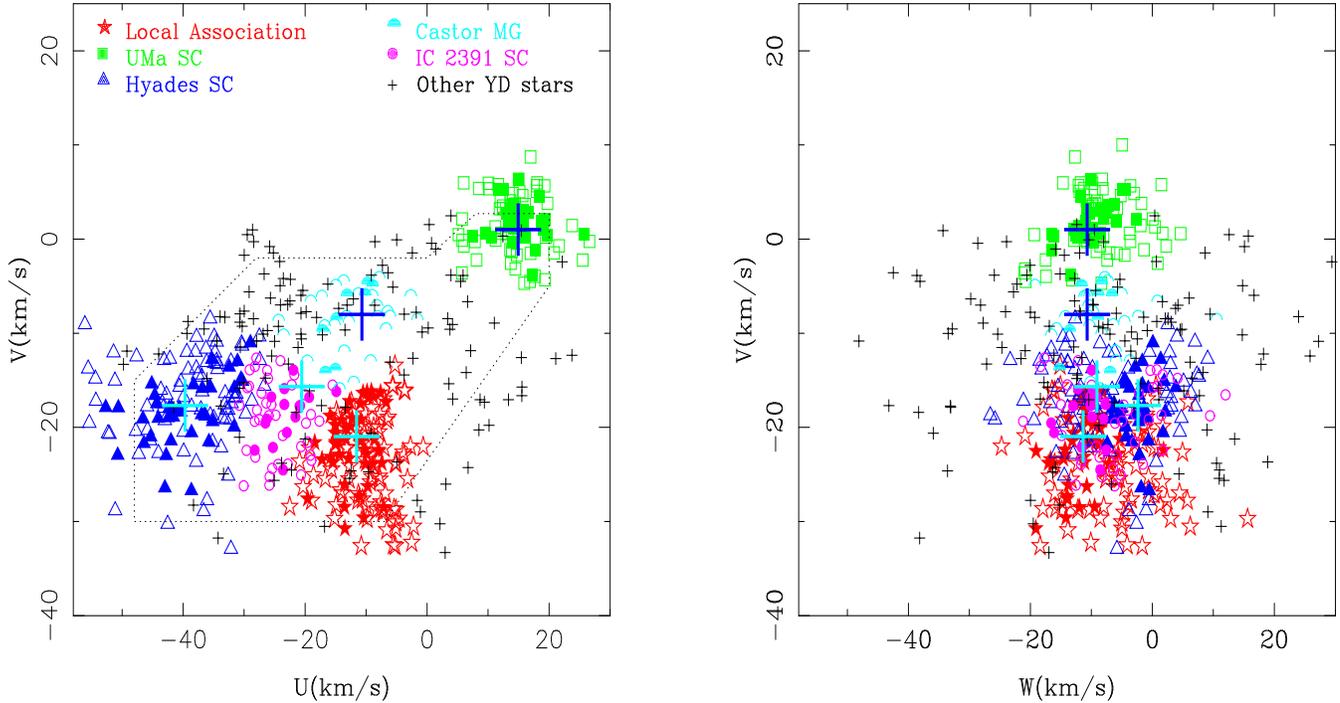}}
\caption[ ]{
(U, V) and (W, V) planes as in Fig.~\ref{fig1}
for our star sample.
We plot with different symbols the stars belonging to the different
stellar kinematic groups, and the other young disk stars.
Filled symbols are stars that satisfied both Eggen's criteria
(peculiar velocity, PV, and radial velocity, $\rho_{c}$),
open symbols are other possible members.
\label{fig2} }
\end{figure*}

\section{Membership to the moving groups}

\subsection{Galactic space-velocity components}

In order to determine the membership of this sample to the different 
SKG we have studied the distribution of stars in 
the velocity space by calculating the 
{\sc Galactic space-velocity components} (U, V, W)
in a right-handed coordinated system (positive in the directions of 
the Galactic center, Galactic rotation, and the
North Galactic Pole, respectively).
We have modified the procedures in Johnson \& Soderblom (1987) 
to calculate U, V, W, and their associated errors.
The original algorithm (which requires epoch 1950 coordinates) is adapted
here to epoch J2000 coordinates 
in the International Celestial Reference System (ICRS) as described in the
Introduction and Guide to the Data (section 1.5) of the 
''The Hipparcos and Tycho Catalogues" (ESA, 1997).
The uncertainties of the velocity components have been obtained using
the full covariance matrix in order to take into account the 
possible correlation between the astrometric parameters.
We have used the correlation coefficients provided by Hipparcos.
It should be noted that the differences, between the errors calculated in this
 way and the obtained considering the covariances are zero 
(as in Johnson \& Soderblom 1987), are very small.
These differences are largely lower than 0.1 km~s$^{-1}$,
only for a small number of star (8) the differences are between 
0.2 and 0.5 km~s$^{-1}$ and only in one case is 1.2 km~s$^{-1}$.
The largest differences are for stars with the largest errors in the input data.

{\sc Parallaxes} and {\sc proper motions} have been taken mainly from
''The Hipparcos and Tycho Catalogues" (ESA 1997) 
and ''The Tycho-2 Catalogue" (H$\o$g et al. 2000), 
which supersedes the
PPM (Positions and Proper Motions) Catalogue 
(R\"{o}ser \& Bastian 1991; Bastian et al., 1993; 
R\"{o}ser, Bastian \& Kuzmin, 1994);
ACT Reference Catalog (Urban, Corbin \& Wycoff 1997); and
TRC (Tycho Reference Catalogue) (H$\o$g et al. 1998).

{\sc Radial velocities} are taken primarily from the compilation
WEB (Wilson Evans Batten) Catalogue (Duflot, Figon \& Meyssonnier 1995),
the Mean radial velocities catalog of galactic stars (Barbier-Brossat \& Figon, 2000) 
which supplements the WEB Catalogue,
the Catalogue of radial velocities of Nearby Stars (Tokovinin 1992),
the Vienna-KPNO search for Doppler-imaging candidate stars, 
 and from other references given in SIMBAD, and in the
 CNS3, Catalogue of Nearby Stars, Preliminary 3rd Version 
(Gliese \& Jahrei$\ss$ 1991) 
or the CNS3R (CNS3 Revised Version)\footnote{ 
CNS3R available only at ARI (Astronomisches Rechen-Institut Heidelberg)
 Database for Nearby Stars at: \\
{\tt http://www.ari.uni-heidelberg.de/aricns/}
}.
For the stars for which we have taken high resolution
echelle spectra (see Montes et al. 2001b, 2001d) 
we have used the radial velocities 
(marked with '*' in Tables 2 to 7) obtained by us 
by cross correlation with radial velocity standard stars
of similar spectral types .  

Our initial sample of more than 1000 stars was reduced to 638 stars with 
accurate parallaxes, proper motions, and radial velocities available 
in the literature to calculate the Galactic space-velocity components (U, V, W).
As we are interested only in young MG we restrict this sample to
the stars which U, V and W components follow the criterion
from Leggett (1992) for young disk stars 
($-50<U<20$; $-30<V<0$; $-25<W<10$) or more exactly to the stars with 
U and V velocity components inside or near the
boundaries (dashed line in Fig.~\ref{fig1} and \ref{fig2}) 
that determine the young disk
population as defined by Eggen (1984b, 1989, 1998a).
We have found 535 stars that satisfied this restriction.

In Tables 2 to 7 we list the stellar and astrometric data we have
compiled for the stars in each SKG.
We give the name (HD, Henry Draper number; variable star name or other name; 
HIP, Hipparcos identifier; GJ, Gliese Catalogue number), 
spectral type, coordinates (ICRS J2000.0), 
radial velocity (V$_{r}$) and the error in km~s$^{-1}$,
parallax ($\pi$) and the error in milli arc second (mas),
proper motions  $\mu$$_{\alpha}$$\cos{\delta}$ and  $\mu$$_{\delta}$
and their errors in mas per year (mas/yr).
The U, V, and W, calculated velocity components with their associated
errors in km~s$^{-1}$ are also given in theses Tables.

In Fig~\ref{fig2} we represent the (U, V) and (W, V) planes 
 for this restricted star sample.
The distribution of the stars in this figure shows concentrations 
around the central (U, V, W) position corresponding to the five MG 
listed in Table~\ref{tab:skg}.
To start the classification, following Eggen's membership criterion 
of constant V, we have considered as members only stars with small
V dispersions.
However, taking into account the results found by other authors 
(Skuljan et al. 1997, 1999)
we have considered a large dispersion in the U, V components 
($\approx$ 8 km~s$^{-1}$) with respect to the central position of the MG 
in the (U, V) plane to classify a star as a possible member. 
In addition, we have also taken into account the information provided by the 
W component, in the sense that stars considered as possible members 
for their position in the (U, V) plane are excluded if their W component
desviates considerably ($\approx$ 8 km~s$^{-1}$)
 with respect to the W component of the MG. 
Following these we have classified the stars from our
sample as members of one of the MG or as other young disk
stars if their classification is not clear but it is inside or near the
boundaries that determine the young disk population (see Tables 2 to 7).
In Fig~\ref{fig2} 
we plot with different symbols the stars belonging to the different
SKG, and the other young disk stars.
Filled symbols represent stars that, in addition, 
satisfied the Eggen's criteria
described in the next two subsections.



\begin{figure*}
{\psfig{figure=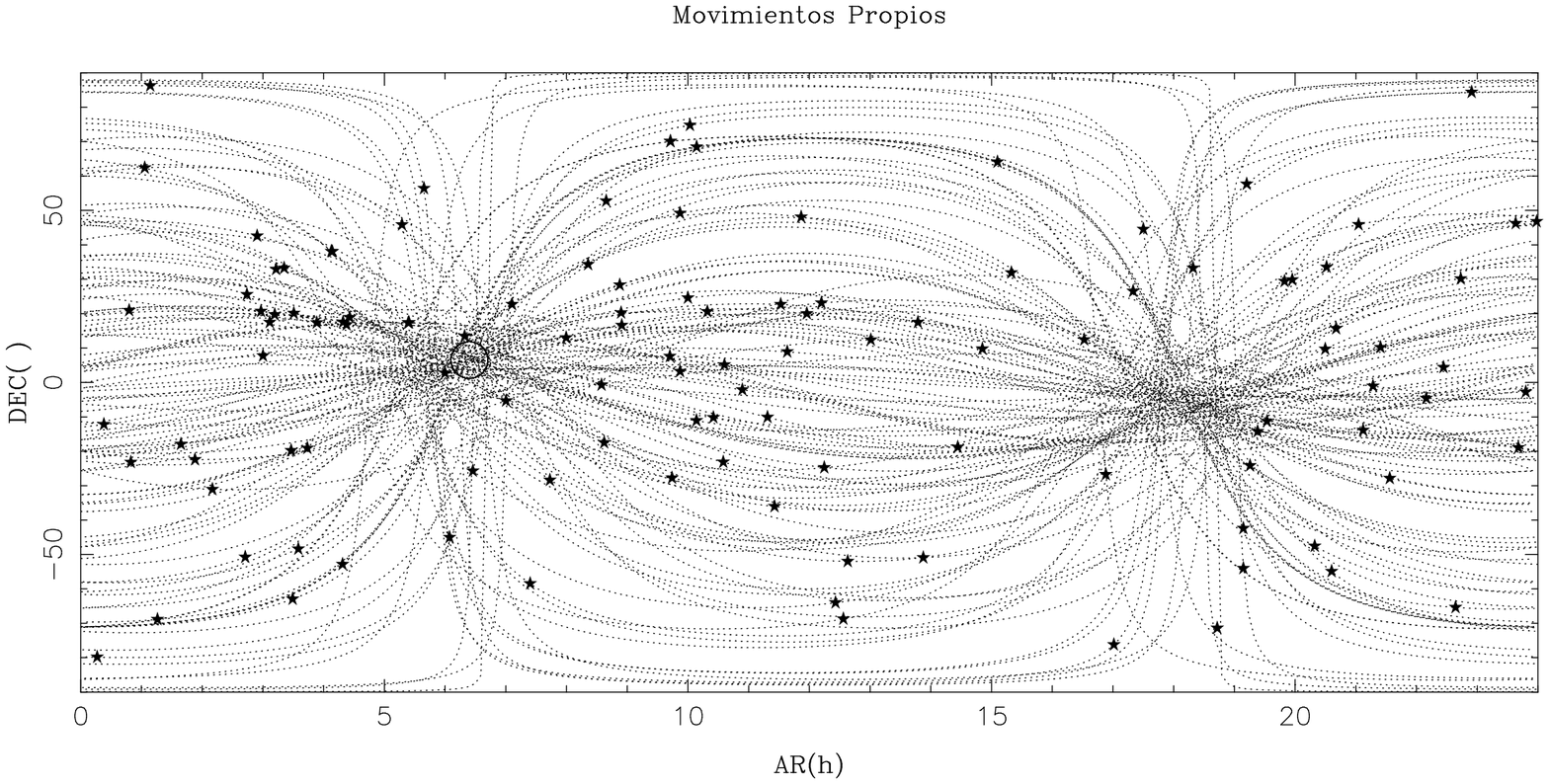,bbllx=38pt,bblly=26pt,bburx=703pt,bbury=345pt,height=9.0cm,width=17.7cm,clip=}}
\caption[ ]{
Spacial position (AR, Dec) for the possible stars members of the Hyades MG.
The convergent point of the MG at AR=6.4$^{h}$, DEC=6.5$^{o}$ 
is marked with a circle.
The dashed lines represent great circles defined by the proper 
motions and position of each star.
\label{fig3} }
\end{figure*}

\subsection{Convergent Point}

The members of a moving group can be established by the degree to
which their motions define a common convergent point in the sky. 
However, this is not a sufficient membership
criterion (there might be some stars moving in the same direction, but
with a significantly different speed).
Eggen's criteria of membership (described in next subsection) 
take also into account the magnitude of the velocity vector.

We can apply the convergent point criterion to a moving group 
by plotting the great circles defined 
by the proper motions and positions of individual stars and analyse if 
their poles are close to the convergent point given in the literature for 
that moving group.
We have applied this analysis to our sample of candidate members to the 
five moving groups studied in this paper 
(see Fig.~\ref{fig3} for the case of the Hyades). 
We have obtained, in general, a good agreement between the position of the 
poles and the convergent point.
However, there are some stars with clear discrepancies, which are probably
not members. 
These deviations with respect to the convergent point will be analysed in a 
more quantitative way by applying the Eggen's criteria as is described 
in next subsection.     

In this convergent point analysis (following other authors) we have 
not corrected for the Sun's motion. It is possible that the Sun's motion induces
this effect so we need to prove that moving groups really converge towards a
point independently of this effect.
In fact, nearly all moving group's convergent points are situated close to
the apex or ant-apex.
If we make this correction with each moving group
we obtain that in the cases of 
the Hyades, Ursa Major, IC 2391 and Castor moving groups 
the trend of the candidate star members
to have a common convergent point, is maintained.
However, in the case of the Local Association the dispersion 
increases somewhat. It  seems that the
proper motion great circles tend to converge towards several points 
that are close together.
This could indicate that the Local Association  has several substructures or
that is a concentration of different moving groups
with similar space motions.

\subsection{Eggen's criteria}

Eggen has developed several criteria during many years studying
stars in moving groups (see Eggen 1958a, 1995b). 
These criteria are based on one supposition: it is possible to
treat moving groups, whose stars are extended in space,
like moving clusters, whose stars are concentrated in space.
As in the moving cluster method it is assumed that the total space
velocities of stars in a moving group are
parallel and move towards a common convergent point. 
The Eggen's criteria try to quantify how the space motion of the stars 
 deviates from the convergent point and
use the following parameters and relations: 

\begin{itemize}

\item  The components 
of the absolute proper motion ($\mu$) in 
the direction of the convergent point ($\upsilon$) 
and perpendicular to it ($\tau$). 

\item The angular distance between the star and the
convergent point ($\lambda$).

\item The trigonometric parallax ($\pi$)

\item The relations between the tangential (V$_{\rm tan}$), 
radial (V$_{\rm r}$) and total (V$_{\rm Total}$) velocities 
in the moving cluster method:

\[
   \rm{V}_{tan} = 4.74 \cdot \mu \cdot \pi^{-1}
\]
\[
   \rm{V}_{tan} = \rm{V}_{Total} \cdot sin \lambda ; \  \  \
   \rm{V}_{r} = \rm{V}_{Total} \cdot cos \lambda
\]
\[
   \rm{V}_{Total} = 4.74 \cdot \mu \cdot \pi^{-1} \cdot sin^{-1} \lambda
\]           
The total velocity can also be calculated from the U, V, W components as:

\[
   \rm{V}_{Total} = (U^{2} + V^{2} + W^{2})^{1/2}
\]

\end{itemize}
The two main Eggen's criteria are:

1) {\sc Peculiar velocity criterion}
\\
In the first papers Eggen used the ratio ($\tau$/$\upsilon$) as a measure
of how the star turns away from the convergent point, but later he defined
a parameter he called Peculiar Velocity (PV) that is defined as a 
V$_{\rm tan}$ but taking into account only the proper motion component 
perpendicular to the C.P. ($\tau$).

\[             
   \rm{PV} = 4.74 \cdot \tau \cdot \pi^{-1}
\]              
The criterion compares this peculiar velocity with another parameter 
he called total velocity (V$_{\rm T}$)  obtained as a real V$_{\rm Total}$ 
but taking into account only the proper motion component in 
the direction of the  C.P. ($\upsilon$).

\[             
    \rm{V}_{T} = 4.74 \cdot \upsilon \cdot \pi^{-1} \cdot sin^{-1} \lambda
\]              
The criterion considers a star as a possible member of a MG when
its peculiar velocity (PV) is less than about 10\% 
of its total velocity (V$_{\rm T}$)

\[ \rm{PV} < 0.1 \cdot  \rm{V}_{T}
\]
Taking into account the definition of  PV and V$_{\rm T}$ this condition can also be
written in terms of components $\tau$ and $\upsilon$ as

\[ \tau/\upsilon < 0.1 \cdot  \sin^{-1}{\lambda} 
\]
This criterion takes into account the information provided by the
proper motion of the star but not the radial velocity.

2) {\sc Radial velocity criterion}
\\
For the moving cluster method we can obtain a predicted radial velocity
(called $\rho_{\rm c}$ by Eggen) as:
\\

\[
   \rho_{\rm c} = \rm{V}_{T} \cdot cos \lambda
\]
The criterion is based in the comparison of this predicted radial velocity
with the observed radial velocity of the star.
Eggen considered a star as a possible member of a MG when
these two velocities differ by less than 4--8 km~s$^{-1}$ 
depending on the quality of the observed radial velocity.

We have applied both criteria (PV, and $\rho_{\rm c}$)  
to our candidate stars (for the five MG), 
in addition to the information provided by the
galactic velocity components (U, V, W, previous section), 
in order to apply more strict requirements for SC membership and
to better discern their membership to the different MG.
For the peculiar velocity criterion we have used the 10\% of V$_{\rm T}$ 
for all the MG except for the Local Association where we have used 
20\% of V$_{\rm T}$, 
to take into account the large dispersion observed in this MG.
For the radial velocity criterion we have taken into account the 
uncertainties of the adopted radial velocity of each star.
In Tables 2 to 6 we list the total velocity (V$_{\rm Total}$) and 
the parameters needed to apply the criteria
(PV, V$_{\rm T}$, and $\rho_{\rm c}$).
The results of applying the PV and $\rho_{\rm c}$ criteria 
are indicated in the column beside each parameter with the labels
"Y" (if possible member) and "N" (if star does not satisfy that criterion).
Errors from V$_{\rm T}$, PV and $\rho_{\rm c}$ are taking into account inside criteria.
In the (U, V) and (W, V) diagrams (Fig.~\ref{fig2}) 
we have plotted with filled symbols,
the stars that satisfied both criteria.

\section{Discussion and Conclusions}

Making use of a great quantity of data from the literature 
(previous general kinematic studies of moving groups, 
many works on late-type stars, 
new results from X-ray surveys, etc.), 
the accurate astrometric data recently released by the
Hipparcos and Tycho-2 catalogues,  
and additional data obtained for our own spectroscopic observations, 
we were able to identify a considerable population of single late-type stars
(for binaries see Paper II) members of young (20--600 Myr) 
stellar kinematic groups.
We have used as membership criteria the position of the stars 
in the velocity space
(U, V, W) and the Eggen's kinematic criteria of deviation of the space motion 
of the star from the convergent point 
and comparison between the observed and calculated radial velocities.
Additional criteria using age-dating methods for late-type stars 
(lithium $\lambda$6708 \AA$\ $ absorption line, 
location on the color-magnitude diagram, and
level of chromospheric and coronal activity)
will be applied in the more detailed study of each SKG 
we have undertaken and will be addressed in forthcoming papers.

In this paper we give the list of possible members (see Tables 2 to 7), 
for each star we list the stellar parameters we have compiled, as well as
the computed galactic space motions and the results of applying the
kinematic criteria.
These data are also available in tabular format and in search-able 
catalogue format in the web page 
{\tt http://www.ucm.es/info/Astrof/skg.html}
that we maintain about stellar kinematic groups. 

For our extensive initial sample of single late-type stars we have found
a total of 535 stars that can be considered, for their position in the
velocity space (U, V, W) as young disk stars.
We have classified 
120 stars as possible members of the Local Association,
118 of the Hyades supercluster,
84  of the Ursa Major moving group,
53  of the IC 2391 supercluster,
34  of the Castor moving group, and 
126 as other young disk stars 
(classification is not clear but it is inside or near the
boundaries that determine the young disk population in the velocity space).

When we take into account the Eggen's kinematic criteria, in the four MG 
where the convergent point is available, 
the number of possible members in each MG is reduced. 
Eliminating only the stars that do not satisfy one of the two criteria 
(peculiar velocity and radial velocity) we found
104 possible members of the Local Association,
96 of the Hyades,
69 of the Ursa Major, 
43 of the IC 2391, and
29 of the Castor.
Considering only the stars that satisfied the peculiar velocity criterion 
we found
77 in the Local Association,
67 in the Hyades, 
37 in the Ursa Major, 
28 in the IC 2391, and
10 of the Castor.
Finally, imposing both criteria the number of possible members is reduced to
45 in the Local Association,
38 in the Hyades, 
28 in the Ursa Major, 
15 in the IC 2391, and
8 of the Castor.

Analyzing these results with the help of the 
great circles defined by the proper motions (see Fig.~\ref{fig3})
we can see that almost all stars which do not
satisfied the PV criterion
move clearly away from the convergent point of the MG, 
specially those which do not
satisfy radial velocity criteria either. 
In the velocity space (U, V, W) the stars which satisfied both criteria
tend to have lower dispersion with respect to the
expected (U, V, W) position of the MG (see Fig.~\ref{fig2}), 
but there are some cases where this
is not true. The latter are normally stars with large errors 
in U, V, and W (due to large errors in radial velocity or
in parallax). 
  

Our results confirm the membership of several
previously established members of SKG,
but in other cases the new calculated Galactic space motions indicate
the membership to a different SKG or that the star should be considered only as
a young disk star with no clear membership to any SKG (e.g. LQ Hya).
Even, in some cases, the new calculations
located the star outside the boundaries of the
young disk population in the Boettlinger Diagram.

For the late-type stars with planetary companions included in our sample
we have found that some stars known to be young: 
GJ 3021 (Naef et al. 2000);
$\iota$ Hor (K\"urster et al. 2000);
$\tau$ Boo (Henry et al. 2000);
55 Cnc (Fuhrmann et al. 1998);
HD 108147 (Mayor et al. 2000),
could be possible members of the Hyades supercluster.
Some of these have been also identified by 
Suchkov \& Schultz (2001) as stars with planetary systems with age similar to
the Hyades.

The groups of nearby late-type stars with different ages 
we have identified in this work
will be very useful for chromospheric activity studies.
High resolution optical spectroscopic observations of these stars
will provide
a  simultaneous analysis of the different optical
chromospheric activity indicators as well as to obtain 
rotation speed, binarity, variability, and kinematics.
With all this information 
it will be possible to study in detail the chromosphere, discriminating between
the different structures: plages, prominences, flares and microflares
(see Montes et al. 2000b, 2001b, 2001d) and analyse the flux-flux 
and rotation-activity relationships and their age evolution.

A further study of the list of stars compiled here
as well as detailed analysis of the origin of these young SKG 
and their relation with nearby young open clusters, OB associations,
T associations, and other recently identified associations of young stars
could lead to a better understanding of the
star formation history in the solar neighbourhood.

Another important usefulness of the list of late-type stars we give here 
is that the youngest ones (the possible members of the Local Association)
can be taken as search targets for direct imaging detection of 
sub-stellar companions (brown dwarfs and extra-solar giant planets).
These young and nearby cool dwarfs favor the optimization 
of the dynamical range and the sub-stellar companions
can be detected directly because they are considerably more luminous when 
undergoing the initial phases of gravitational contraction 
than at later stages. Until now only five brown dwarfs have been 
detected directly (and confirmed by both spectroscopy and
proper motion) as companions to nearby stars, 
the T dwarf Gl 229~B (Nakajima et al. 1995),  
the young L dwarf G 196-3~B (Rebolo et al. 1998),
the T dwarf Gl 570~D (Burgasser et al. 2000),
the M9 dwarf CoD-33$^{\circ}$~7795~B (Neuh\"auser et al. 2000b) 
which is a member of the TW Hya association, and
the M8 dwarf HR 7329~B (Guenther et al. 2001) 
which is a member of the Tucanae association.
The B component of the Ursa Major group member Gl 569
 seem to be a triple brown dwarf system (Mart\'{\i}n et al. 2000; 
Kenworthy et al. 2001).
In addition, Neuh\"auser et al. (2000a) have shown 
that direct imaging detection of
extra-solar giant planets is already possible with current technology.

Radial velocity is an important parameter in the determination of the 
space velocity components, and in some cases only poor quality measurements
are available in the literature, resulting in large errors in U, V, and W.
Good quality spectroscopic observations are needed to confirm 
the membership of these stars to a SKG. 
We have already started a program of high resolution
echelle spectroscopic observations (using 2m class telescopes) 
of these candidate stars in order to obtain
a better determination of their radial velocity, as well as other 
stellar parameters.
We will use these new data to better establish the membership 
of these stars (for preliminary results see  Montes 2001b;
Montes et al. 2001b, 2001d).  
 
However, a considerable number of stars in our initial sample, are too faint 
and no radial velocities or accurate astrometric parameters 
are available in the literature.
High resolution spectroscopic observations using 4 or 8m class telescopes  
will be needed to obtain the spectroscopic parameters of these stars.
Accurate astrometric parameters for a huge number of stars 
will be available, in the future, with the space-astrometry missions
 DIVA (Double Interferometer for Visual Astrometry) and
 FAME (Full-sky Astrometric Mapping Explorer).
The space mission  GAIA (Global Astrometric Interferometer for Astrophysics)   
will have a much large reach in distance (magnitude limit 20) and 
will provide both astrometric data and radial velocities.

\section*{Acknowledgments}

This research has made use the 
of the SIMBAD database, operated at CDS,
Strasbourg, France, and the ARI Database for Nearby Stars, 
Astronomisches Rechen-Institut, Heidelberg.
We would like to thank Dr. D. Barrado y Navascu\'es for provided us 
some additional candidates to our initial sample of late-type stars.
We would also like to thank the anonymous referee for suggesting several
improvements and helpful comments.
This work was supported by the Universidad Complutense de Madrid and
the Spanish Direcci\'{o}n General de Ense\~{n}anza Superior e 
Investigaci\'{o}n Cient\'{\i}fica (DGESIC) under grant PB97-0259.





\begin{thebibliography}{}
%
\bibitem{} Agekyan T.A., Orlov V.V., 1984, Astron. Zh., 61, 60, (SvA 28, 36)
%
\bibitem{} Ambruster C.W., Brown A., Fekel F.C., Harper G.H., Fabian D.,
Wood B., Guinan E.F., 1998, in ASP Conf. Ser. 154, The Tenth Cambridge
Workshop on Cool Stars, Stellar Systems, and the
Sun, eds. R.A. Donahue, J.A. Bookbinder,
CD-1205

\bibitem{} Asiain R., Figueras F., Torra J.,  Chen B., 1999, A\&A, 341, 427

\bibitem{} Barbier-Brossat M., Figon P., 2000, A\&AS, 142, 217)

\bibitem{} Barrado y Navascu\'es D., 1998, A\&A, 339, 831

\bibitem{} Barrado y Navacu\'es D., Stauffer J.R., Song I., 
Caillault J.P., 1999, ApJ, 520, L123

\bibitem{} Bastian U., et al. 1993,  
Astronomisches Rechen- Institut, Heidelberg. 
Spektrum Akademischer Verlag, Heidelberg

\bibitem{} Bochanski J.J., et al.,
2000, AAS, 196, 4607

\bibitem{} Brown A.G.A., 2001, ASP Conf. Ser., in:
Modes of Star Formation and the Origin of Field Populations, 
eds. E.K. Grebel, W. Brandner

\bibitem{} Burgasser A.J., et al., 2000, ApJ, 531, L57

\bibitem{} Chereul E., Cr\'ez\'e M., Bienaym\'e O., 1998, A\&A, 340, 384

\bibitem{} Chereul E., Cr\'ez\'e M., Bienaym\'e O., 1999, A\&AS, 135, 5

\bibitem{} Chugainov P.F., 1991, in Angular Momentum Evolution of Young Stars,
S. Catalano, J.R. Stauffer (eds.), Kluwer Acad. Publ., p. 175

\bibitem{} Cutispoto G., Pastori L., Tagliaferri G., Messina S., 
Pallavicini R., 
1999, A\&AS, 138, 87

\bibitem{} Cutispoto G., Pastori L., Guerrero A., Tagliaferri G., Messina S., 
Rodon\`o M., de Medeiros J.R.,
2000, A\&A, 364, 205

\bibitem{} Dehnen W., 1998, AJ, 115, 2384

\bibitem{} Delfosse X., Forveille T., Perrier C.,  Mayor M.,
1998, A\&A, 331, 581

\bibitem{} Delfosse X., et al.,
2001, A\&A, 366, L13

\bibitem{} Duflot M., Figon P., Meyssonnier N., 1995 A\&AS, 114, 269

\bibitem{} Eggen O.J., 1958a, MNRAS, 118, 65

\bibitem{} Eggen O.J., 1958b, MNRAS, 118, 154





\bibitem{} Eggen O.J., 1960a, MNRAS, 120, 540

\bibitem{} Eggen O.J., 1960b, MNRAS, 120, 563


\bibitem{} Eggen O.J., 1975, PASP, 87, 37

\bibitem{} Eggen O.J., 1983a, AJ, 88, 642

\bibitem{} Eggen O.J., 1983b, MNRAS, 204, 377

\bibitem{} Eggen O.J., 1983c, MNRAS, 204, 391

\bibitem{} Eggen O.J., 1984a, AJ, 89, 1358

\bibitem{} Eggen O.J., 1984b, ApJS, 55, 597

\bibitem{} Eggen O.J., 1984c, AJ, 89, 1350

\bibitem{} Eggen O.J., 1989, PASP, 101, 366

\bibitem{} Eggen O.J., 1991, AJ, 102, 2028

\bibitem{} Eggen O.J., 1992a, AJ, 104, 1493

\bibitem{} Eggen O.J., 1992b, AJ, 104, 1482

\bibitem{} Eggen O.J., 1992c, AJ, 103, 1302

\bibitem{} Eggen O.J., 1994, L.V. Morrison, G. Gilmore (eds.),
      Galactic and Solar System Optical Astrometry,
      Cambridge University Press, 191

\bibitem{} Eggen O.J., 1995a, AJ, 110, 1749

\bibitem{} Eggen O.J., 1995b, AJ, 110, 2862

\bibitem{} Eggen O.J., 1996, AJ, 111, 1615

\bibitem{} Eggen O.J., 1998a, AJ, 115, 2397

\bibitem{} Eggen O.J., 1998b, AJ, 116, 284

\bibitem{} Eggen O.J., 1998c, AJ, 116, 782

\bibitem{} ESA, 1997, The Hipparcos and Tycho Catalogues, ESA SP-1200

\bibitem{} Favata F., Barbera M., Micela G., Sciortino S.,
1993, A\&A, 277, 428

\bibitem{} Favata F., Barbera M., Micela G., Sciortino S.,
1995, A\&A, 295, 147

\bibitem{} Favata F., Micela G., Sciortino S., D'Antona F.,
1998, A\&A, 335, 218

\bibitem{} Feltzing S., Holmberg J., 2000, A\&A, 357, 153

\bibitem{} Ford E.B., Rasio F.A., Sills A., 
1999, ApJ, 514, 411

\bibitem{} Frink S., 2001, ASP Conf. S., in: 
Dynamics of Star Clusters and the Milky Way, 
eds. S. Deiters, B. Fuchs, A. Just, R. Spurzem, R. Wielen (in press)

\bibitem{} Fuhrmann K., Pfeiffer M.J., Bernkopf J.,             
1998, A\&A, 336, 942 

\bibitem{} Gaidos E.J., 1998, PASP, 110, 1259

\bibitem{} Gaidos E.J., Henry G.W., Henry S.M., 2000, AJ, 120, 1006

\bibitem{GJ1991} Gliese W., Jahrei$\ss$ H., 1991,
 Preliminary Version of the Third Catalogue of Nearby Stars,
 (Astron. Rechen-Institut, Heidelberg) (CNS3)


\bibitem{} Griffin R.F., 1998, The Observatory, 118, 223

\bibitem{} Guenther E.W., Neuh\"auser R., Hu\'elamo N., Brandner W., Alves J.,
2001, A\&A, 365, 514

\bibitem{} Guillout P., Sterzik M.F., Schmitt J.H.M.M., Motch C.,
 Neuh\"auser R.,
1998, A\&A, 337, 113

\bibitem{} Henry G.W., Fekel F.C., Hall D., 1995, AJ, 110, 2926

\bibitem{} Henry G.W., Baliunas S.L., Donahue R.A., Fekel F.C., Soon W.,
2000, ApJ, 531, 415

\bibitem{} Henry T.J., Soderblom D.R., Donahue R.A., Baliunas S.L.,
1996, AJ, 111, 439

\bibitem{} H$\o$g E., et al., 1998, A\&A, 335, L65

\bibitem{} H$\o$g E., et al., 2000, A\&A, 355, L27

\bibitem{} H\"unsch M., Schmitt J.H.M.M., Voges W., 
1998a, A\&AS, 127, 251 

\bibitem{} H\"unsch M., Schmitt J.H.M.M., Voges W.,
1998b, A\&AS, 132, 155

\bibitem{} H\"unsch M., Schmitt J.H.M.M., Sterzik M.F., Voges W.,
1999, A\&AS, 135, 319

\bibitem{} Jeffries R.D., Jewell S.J., 1993, MNRAS, 264, 106

\bibitem{} Jeffries R.D., 1995, MNRAS, 273, 559

\bibitem{} Johnson D.R.H., Soderblom D.R., 1987, AJ, 93, 864

\bibitem{} Kenworthy M.A., et al., 2001, ApJ, 554, L67

\bibitem{} K\"urster M., Endl M., Els S., Hatzes A.P., Cochran W.D., 
D\"obereiner S., Dennerl K.,
2000, A\&A, 353, L.33

\bibitem{} Leggett S.K., 1992, ApJS, 82, 351

\bibitem[2001]{LMG01} L\'opez-Santiago J., Montes D., G\'alvez M.C.,
 2001, Ap\&SS, in:
 Proceedings of the IV Scientific Meeting of the Spanish Astronomical Society
(SEA), eds. J. Zamorano, J. Gorgas, J. Gallego (in press)

\bibitem{} Makarov V.V., Urban S., 2000, MNRAS, 317, 289

\bibitem{} Mart\'{\i}n E.L., Magazz\`u A., 1999 A\&A 342, 173 

\bibitem{} Mart\'{\i}n E.L., Koresko C.D., Kulkarni S.R., Lane B.F., 
Wizinowich P.L., 
2000, ApJ 529, L37

\bibitem{} Marcy G.W., Cochran W.D., Mayor M.,
2000, Protostars and Planets IV 
(Book - Tucson: University of Arizona Press; 
eds. Mannings V., Boss A.P., Russell S.S.), p. 1285

\bibitem{} Marcy G.W., Butler R.P.,
2000, PASP, 112, 137

\bibitem{} Mayor M., Naef D., Pepe F., Queloz D., Santos N.C.,
 Udry S., Burnet M., 2000, ESO Press Release 13/00
(http://obswww.unige.ch/$\sim$udry/planet/hd108147.html)

\bibitem[2001a]{M01a} Montes D.,
 2001a, ASP Conf. Ser. 223, CD-1471, in: 
The 11th Cambridge Workshop on Cool Stars,
Stellar Systems, and the Sun,
eds. R. Garc\'{\i}a L\'{o}pez, R. Rebolo, M.R. Zapatero Osorio 

\bibitem[2001b]{M01b} Montes D.,
 2001b, Ap\&SS, in: 
 Proceedings of the IV Scientific Meeting of the Spanish Astronomical Society 
(SEA), eds. J. Zamorano, J. Gorgas, J. Gallego (in press)

\bibitem[2000]{MLF00a} Montes D., Latorre A., Fern\'andez-Figueroa M.J.,
2000a, ASP Conf. Ser. 198, p. 203, in: 
Stellar clusters and associations: convection,
rotation, and dynamos, eds. R. Pallavicini, G. Micela,  S. Sciortino.

\bibitem[2000]{M_etal00b} Montes D., Fern\'andez-Figueroa M.J., De Castro E.,
Cornide M., Latorre A., Sanz-Forcada J., 
2000b, A\&AS 146, 103

\bibitem[2001a]{M_etal01a} Montes D., Fern\'andez-Figueroa M.J., De Castro E.,
Cornide M., Latorre A.,
 2001a, ASP Conf. Ser. 223, CD-1477, in: 
The 11th Cambridge Workshop on Cool Stars,
Stellar Systems, and the Sun,
eds. R. Garc\'{\i}a L\'{o}pez, R. Rebolo, M.R. Zapatero Osorio 

\bibitem[2001b]{MLG01b}  Montes D., L\'opez-Santiago J., G\'alvez M.C.,
 2001b, Ap\&SS, in:
 Proceedings of the IV Scientific Meeting of the Spanish Astronomical Society
(SEA), eds. J. Zamorano, J. Gorgas, J. Gallego (in press)

\bibitem[2001c]{PaperII}  Montes D., G\'alvez M.C., L\'opez-Santiago J.,  
Fern\'andez-Figueroa M.J.,  Cornide M., De Castro E. 
2001c, MNRAS, (submitted) (Paper II)

\bibitem[2001d]{}  Montes D., L\'opez-Santiago J.,  
Fern\'andez-Figueroa M.J., G\'alvez M.C., 2001d, A\&A, (submitted)

\bibitem{} Myll\"ari A., Flynn C., Orlov V.,
2000, Astronomische Gesellschaft Meeting, 16, T12

\bibitem{} Mullis C.L., Bopp W.B., 1994, PASP, 106, 822

\bibitem{} Naef D., Mayor M., Pepe F., Queloz D., Udry S.,  Burnet M., 
2000, ASP Conf S. 219, in: "Disks, Planetesimals and Planets",
eds. F. Garz\'on, C. Eiroa, D. de Winter, and T. J. Mahoney

\bibitem{} Nakajima T., Oppenheimer B.R., Kulkarni S.R.,  
Golimowski D.A., Matthews K., Durrance S.T.,
1995, Nature, 378, 463

\bibitem{}  Neuh\"auser R., Brandner W., Eckart A., Guenther E., 
Alves J., Ott T., Hu\'elamo N., Fern\'andez M.,
2000a, A\&A, 354, L9

\bibitem{}  Neuh\"auser R., Guenther E., Petr M.G., Brandner W., 
Hu\'elamo N., Alves J.,
2000b, A\&A, 360, L39

\bibitem{} Oppenheimer B.R., Basri G., Nakajima T., Kulkarni S.R.,
1997, AJ, 113, 296

\bibitem{} Orlov V.V., Panchenko I.E., Rastorguev A.S., Yatsevich A.V.,
1995, Astron. Zh., 72, 495

\bibitem{} Popovi\'c G.M., Ninkovi\'c S., Pavlovi\'c R.,
1995, Bull. Astron. de Belgrade, 152, 59

\bibitem{} Poveda A., Allen C., Herrera M.A., Cordero G., Lavalley C., 
 1996, A\&A, 308, 55

\bibitem{} Rebolo R., Zapatero Osorio M.R., Madruga S., Bejar V.J.S., 
Arribas S., Licandro J., 
1998, Science, 282, 1309  

\bibitem{} R\"{o}ser S., Bastian U., 1991, Astronomisches Rechen-Institut,
    Heidelberg. Spektrum Akademischer Verlag, Heidelberg

\bibitem{} R\"{o}ser S., Bastian U., Kuzmin A., 1994, A\&AS, 105, 301

\bibitem{} Schachter J.F., Remillard R., Saar S.H., Favata F., Sciortino S.,
Barbera M., 1996, ApJ, 463, 747

\bibitem{} Skuljan J., Cottrell P.L.,  Hearnshaw J.B., 1997, 
Proceedings of the ESA Symposium `Hipparcos - Venice '97', 
ESA SP-402, 525 

\bibitem{}  Skuljan J., Hearnshaw J.B., Cottrell P.L., 
1999, MNRAS, 308, 731 

\bibitem{} Soderblom D.R., Clements S.D., 1987, AJ, 93, 920

\bibitem{} Soderblom D.R., 1990, AJ, 100, 204

\bibitem{} Soderblom D.R., Mayor M., 1993a, AJ, 105, 226

\bibitem{} Soderblom D.R., Mayor M., 1993b, ApJ, 402, L5

\bibitem{} Soderblom D.R., King J.R., Hanson R.B., Jones B.F., 
Fischer D., Stauffer J.R., Pinsonneault M.H., 1998, ApJ, 504, 192

\bibitem{} Song I., Caillault J.P, Barrado y Navacu\'es D.,
 Stauffer J.R., Randich S., 2000, ApJ, 532, L41




\bibitem{} Strassmeier K.G., Washuettl A., Granzer Th., Scheck M., 
 Weber M., 2000, A\&AS, 142, 275

\bibitem{} Suchkov A.A., Schultz A., 2001, AAS, 197, 4901

\bibitem{} Tagliaferri G., Cutispoto G., Pallavicini R., Randich S.,
Pasquini L., 1994, A\&A, 285, 272

\bibitem{} Taylor B.J., 2000, A\&A, 362, 563

\bibitem{} Tokovinin A.A., 1992, A\&A, 256, 121

\bibitem{} Torra J., Fern\'andez D., Figueras F., 2000, A\&A, 359, 82

\bibitem{} Torres C.A.O., da Silva L., Quast G.R., de la Reza R., Jilinski E.,
 2000, AJ, 120, 1410 

\bibitem{} Upgren A.R., 1988, PASP, 100, 251

\bibitem{} Urban S.E., Corbin T.E., Wycoff G.L., 1997,
  U.S. Naval Observatory, Washington D.C.

\bibitem{} van den Ancker M.E., P\'erez M.R., de Winter D., McCollum B.,
2000, A\&A, 363, L25

\bibitem{} van den Ancker M.E., P\'erez M.R., de Winter D., 
2001, ASP Conf. Series,
Proc. "Young Stars Near Earth: Progress and Prospects", 
eds. R. Jayawardhana \& T. Greene

\bibitem{} Wichmann R., Schmitt J.H.M.M.,  
2001, ASP Conf. Ser., 223, CD-552, in: 
The 11th Cambridge Workshop on Cool Stars,
Stellar Systems, and the Sun,
eds. R. Garc\'{\i}a L\'{o}pez, R. Rebolo, M.R. Zapatero Osorio 

\bibitem{} Wichmann R., Covino E., Alcal\'a J.M., Krautter J., Allain S., 
Hauschildt P.H., 1999, MNRAS, 307, 909

\bibitem{} Wichmann R., et al., 2000, A\&A, 359, 181

\bibitem{} Young A., Sadjadi S., Harlan E.,
1987, ApJ, 314, 272

\bibitem{} Zuckerman B., Webb R.A., 2000, ApJ, 535, 959

%
\end{thebibliography}
\end{document}